\documentclass{appolb}
\usepackage{graphicx}

\begin{document}
\title{Searching for the critical point of strongly interacting matter in nucleus-nucleus collisions at CERN SPS%
\thanks{Presented at the 45th Congress of Polish Physicists, 13 - 18 September 2019, Krakow, Poland}%
}
\author{N.~Davis${}^*$ for the NA61/SHINE Collaboration\\%
\address{${}^*$H.~Niewodnicza\'nski Institute of Nuclear Physics, Polish Academy of
Sciences,\\ul. Radzikowskiego 152, 31-342 Krak\'ow, Poland\\%
	}
	    }
\maketitle
\begin{abstract}
The search for experimental signatures of the critical point (CP) of strongly interacting matter is one of the main objectives of the\linebreak NA61/SHINE experiment at CERN SPS. In the course of the experiment, a beam momentum and system size scan is performed. Local proton density fluctuations in transverse momentum space represent an order parameter of the chiral phase transition and are expected to scale according to a universal power-law in the vicinity of the CP; we probe their behavior through an intermittency analysis of the proton second scaled factorial moments (SSFMs) in transverse momentum space. Previous such analyses revealed power-law behavior in NA49 Si+Si collisions at 158$A$~GeV/$c$, with no intermittency observed in lighter or heavier NA49 \& NA61/SHINE systems at the same energy. We now extend the analysis to NA61/SHINE Ar+Sc collisions at 150$A$~GeV/$c$, similar in size and baryochemical potential to NA49 Si+Si. We employ statistical techniques to subtract non-critical background and estimate statistical and systematic uncertainties. Subsequently, we use Monte Carlo simulations to assess the statistical significance of the observed intermittency effect.
\end{abstract}

  
\section{Introduction}
\label{sec:intro}

NA61/SHINE \cite{SHINE} at the CERN Super Proton Synchrotron (SPS) is a fixed-target experiment, colliding a variety of beams on hydrogen and nuclear targets.

\begin{figure}[htb]
\centering
\includegraphics[width=0.27\textwidth]{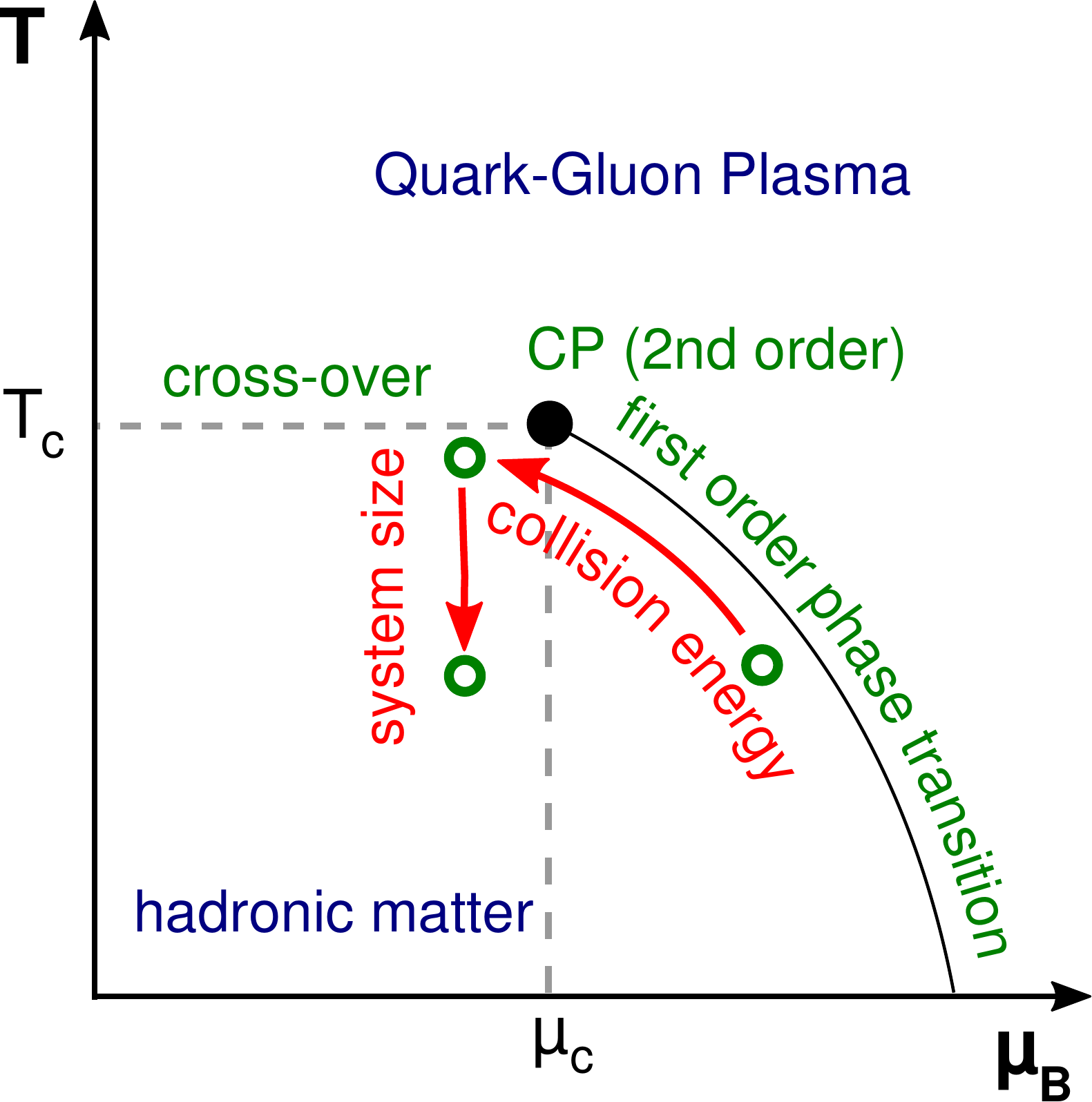}
\includegraphics[width=0.37\textwidth]{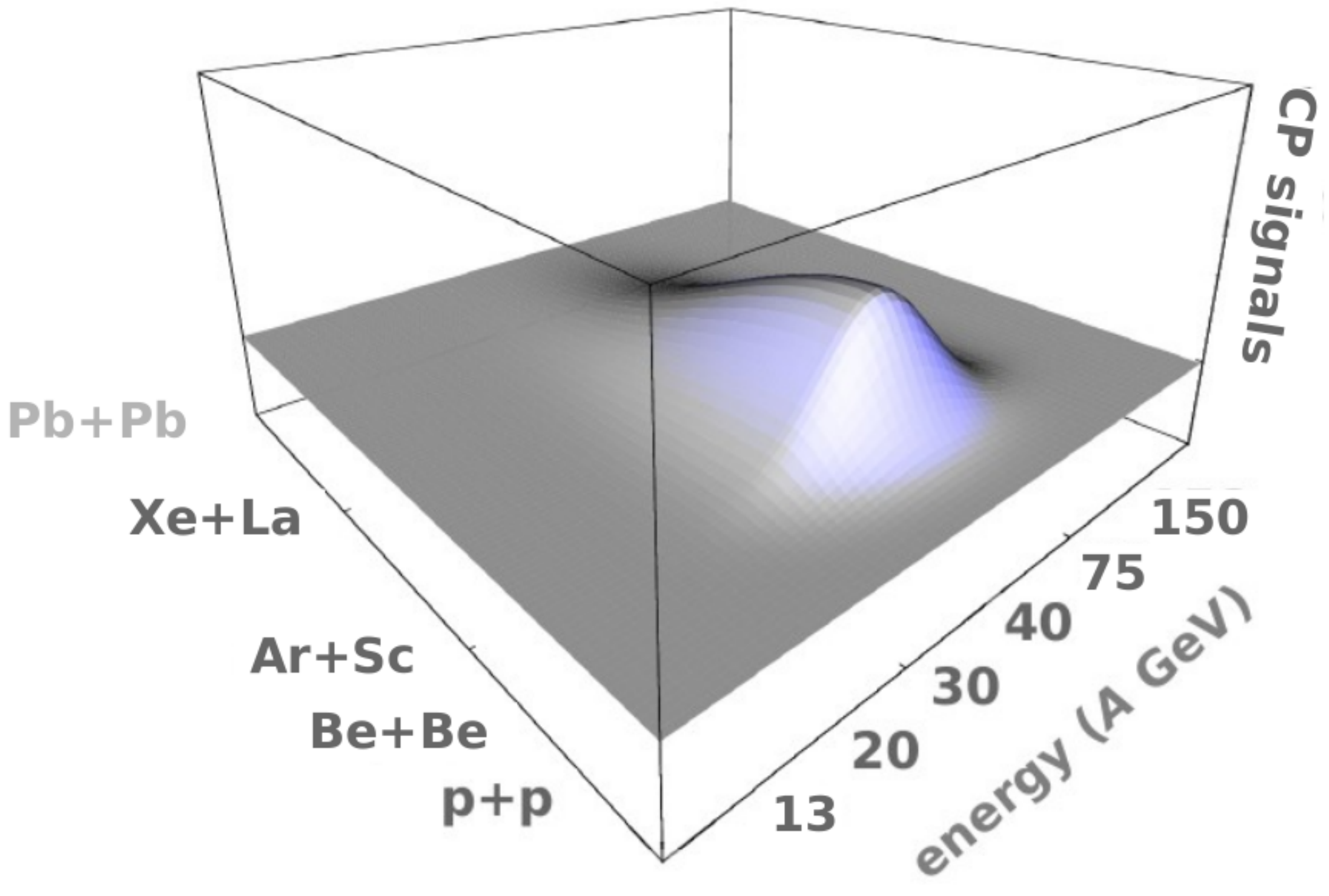}
\caption{{\em Left}: Hypothetical sketch of the phase diagram of strongly interacting matter with critical point, drawn as a function of baryochemical potential $\mu_B$ and temperature $T$. {\em Right}: Theoretical studies predict the presence of a ``hill of fluctuations'' as a function of colliding system size and energy, for observables sensitive to the CP (see Ref.\cite{Gazdzicki} for more details).}
\label{fig:hillscan}
\end{figure}

One of the stated goals of NA61/SHINE is the search for the critical point (CP) of strongly interacting matter. NA61/SHINE is the first experiment to perform a two-dimensional scan, in beam momentum\linebreak (13A -- 150$A$~GeV/$c$) and system size (p+p, p+Pb, Be+Be, Ar+Sc, Xe+La) of colliding nuclei, thus probing different freeze-out conditions in temperature $T$ and baryochemical potential $\mu_B$ (Fig.\ref{fig:hillscan}~left).

Near the CP, a second order phase transition occurs; the correlation length of the system diverges, leading to a scale-invariant system and an expected ``hill'' of increased fluctuations in various observables in the CP vicinity (Fig.\ref{fig:hillscan}~right). Of particular interest are local power-law fluctuations of the net baryon density \cite{Antoniou2005-6}, connected to the order parameter of the QCD chiral phase transition, the chiral condensate. At finite baryochemical potential, critical fluctuations are also transferred to the net proton density, as well as to the proton and antiproton densities separately \cite{Antoniou2005-6,Hatta2003}. At the CP, the fluctuations of the order parameter are self-similar \cite{Vicsek}, belonging to the 3D-Ising universality class, and can be detected in transverse momentum space within the framework of a proton intermittency analysis \cite{Antoniou2005-6,Wosiek-Bialas} by use of scaled factorial moments (SFMs). A detailed analysis can be found in Ref.~\cite{intermittency}, where we study various heavy nuclei collision datasets recorded in the NA49 experiment at maximum energy (158$A$~GeV/$c$, $\sqrt{s_{NN}}\approx 17$~GeV) of the SPS (CERN).

\section{Method of intermittency analysis}
Intermittency is defined as power-law scaling of the Second Scaled Factorial Moments (SSFMs) of protons as a function of bin size in transverse momentum space. The SSFMs are calculated by partitioning a region of transverse momentum space into a lattice of $M \times M$ equal-size bins, and counting the number of proton pairs per bin:

\begin{equation}
F_2(M)=  \left\langle \displaystyle{\frac{1}{M^2}\sum_{i=1}^{M^2}} n_i(n_i-1) \right\rangle  \Bigg/ \left\langle \displaystyle{\frac{1}{M^2}\sum_{i=1}^{M^2}} n_i \right\rangle^2 
\label{eq:facmom}
\end{equation}
 where $n_i$ is the number of particles in the $i$-th bin and $M^2$ is the total number of bins, and we average over bins and events ($\langle\ldots\rangle$). In the case of a pure system exhibiting critical fluctuations, $F_2(M)$ is expected to scale with $M$, for large values of $M$, as a power-law:

\begin{equation}
F_2(M) \sim M^{2\phi_2}, \; \phi_2 = \phi^B_{2,cr}= {}^5/{}_6
\label{eq:phi2}
\end{equation}
where $\phi_2$ is the intermittency index, and provided the freeze-out occurs exactly at the critical point \cite{Antoniou2005-6}.

Noisy experimental data require the subtraction of a background of uncorrelated \& misidentified protons, which is achieved through the construction of correlation-free mixed events. A correlator $\Delta F_2(M)$ can then be defined in terms of the moments of data and mixed events. In the special case where the background dominates over the critical component, Monte Carlo simulations indicate we can approximate the correlator as:

\begin{equation}
 \Delta F_2^{(e)}(M) \simeq F_2^{(d)}(M) - F_2^{(m)}(M),
 \label{eq:correlator}
\end{equation}
where mixed event $(m)$ moments are simply subtracted from data $(d)$ moments \cite{intermittency}. $\Delta F_2(M)$ should then scale as a power law, $\Delta F_2(M) \sim M^{2\phi_2}$, in a limited range, with the same intermittency index as the pure critical system.

SSFMs statistical errors are estimated by the bootstrap method \cite{Metzger, Efron-Hesterberg}, whereby the original set of events is resampled with replacement \cite{intermittency}. Fitting $\Delta F_2^{(e)}(M)$ to obtain $\phi_2$ confidence intervals is complicated by bin correlations among $M$-values. The matter is under current investigation.

A proton generating modification of the Critical Monte Carlo (CMC) code \cite{Antoniou2005-6} is used to simulate a system of critically correlated protons, which are mixed with a non-critical background to study the effects on the quality of intermittency analysis.

\section{Results}

\begin{figure}[htb]
\centering
\includegraphics[width=0.32\textwidth]{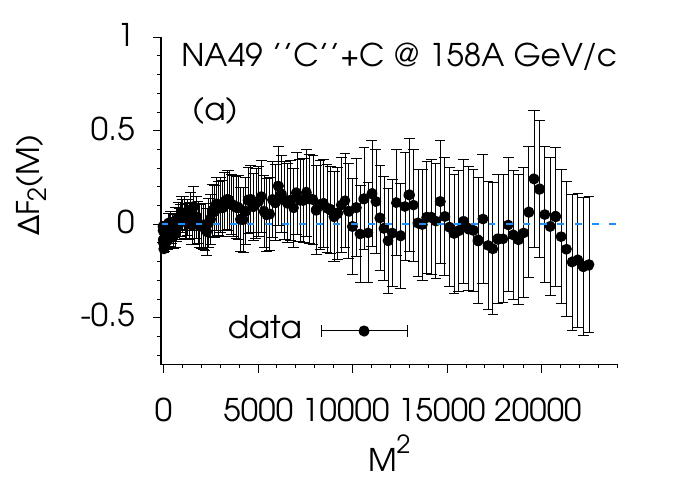}
\includegraphics[width=0.32\textwidth]{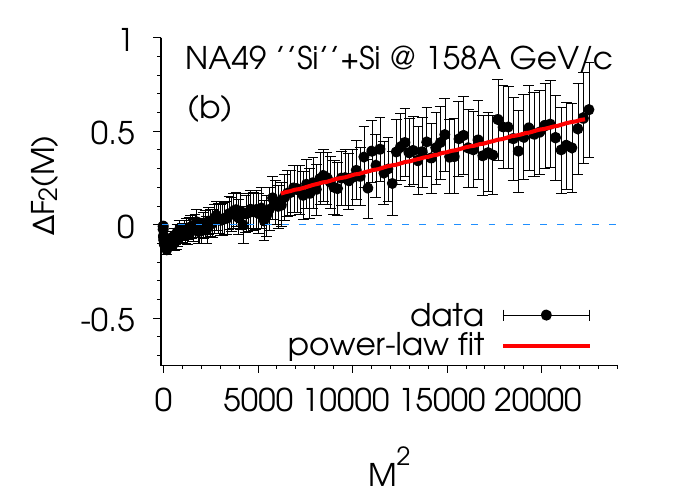}
\includegraphics[width=0.32\textwidth]{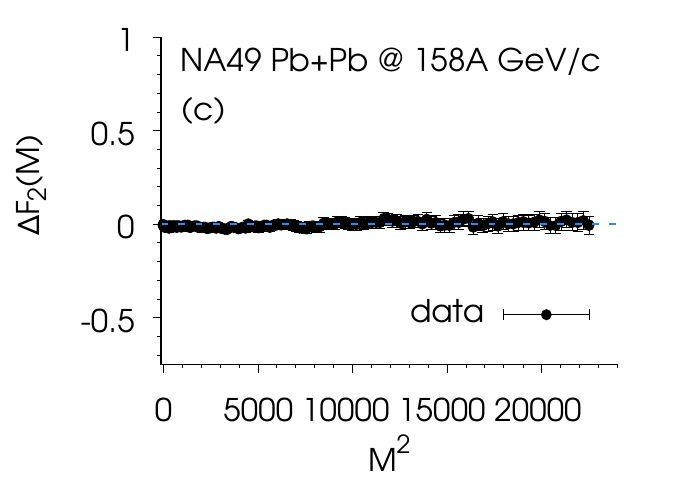}
\caption{$\Delta F_2(M)$ for NA49 {\em (a)} C+C and {\em (b)} Si+Si (0-12\% most central), and {\em (c)} Pb+Pb collisions (0-10\% most central) at 158$A$~GeV/$c$  (Ref.\cite{intermittency}).}
\label{fig:DF2_experiments_NA49}
\end{figure}

Proton intermittency analysis was first performed on data collected by the NA49 experiment \cite{intermittency}. Three collisions systems of different size were analyzed: C+C, Si+Si and Pb+Pb at mid-rapidity, at the maximum SPS energy of 158$A$~GeV/$c$. Fig.\ref{fig:DF2_experiments_NA49}(a-c) shows the correlator $\Delta F_2(M)$ as a function of bin size $M$ for the analyzed systems. No intermittency was detected in C+C and Pb+Pb; by contrast, the Si+Si system exhibits power-law fluctuations compatible with criticality, with an intermittency index value estimated, through the bootstrap, as $\phi_{2,B} = 0.96_{-0.25}^{+0.38}(\mathrm{stat.})\pm 0.16(\mathrm{syst.})$~\cite{intermittency}.

\begin{figure}[htb]
\centering
\includegraphics[width=\textwidth]{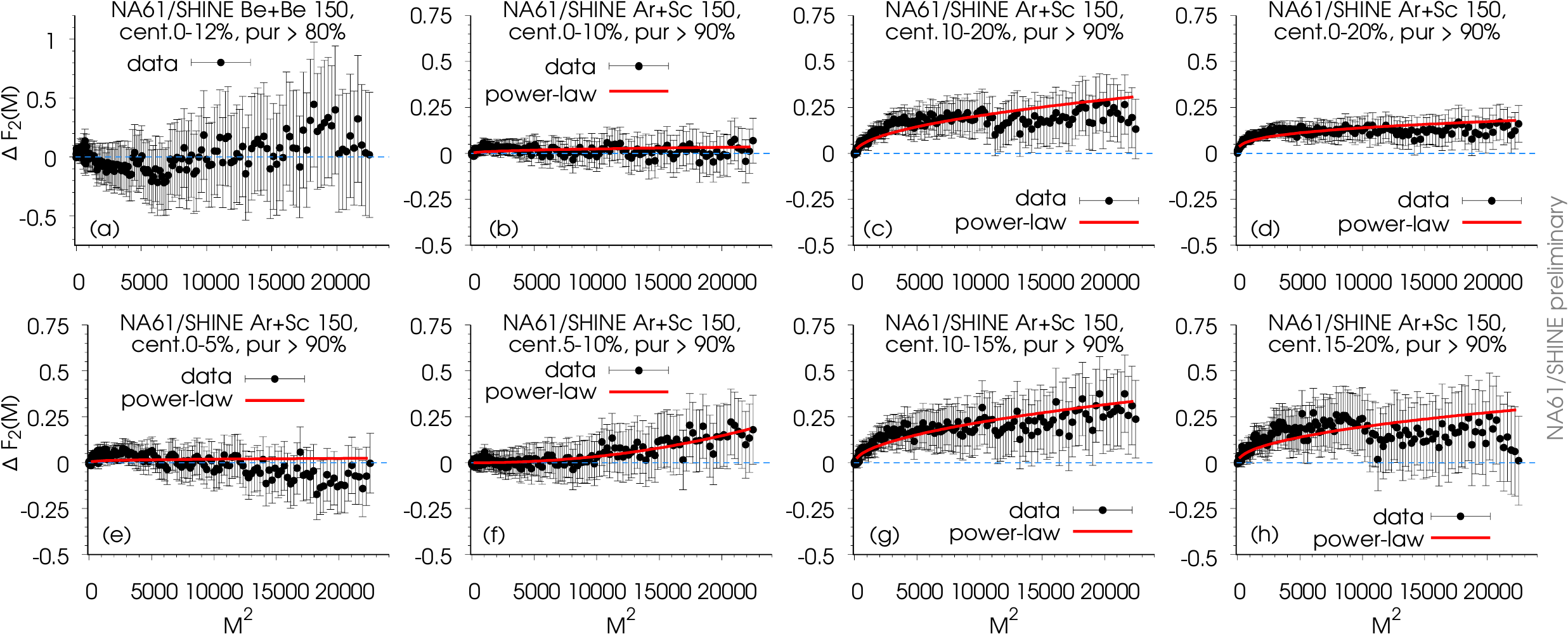}
\caption{$\Delta F_2(M)$ for NA61/SHINE {\em (a)} Be+Be (0-12\% most central) collisions \cite{PoS2017}, as well as Ar+Sc collisions at {\em (b)} 0-10\%, {\em (c)} 10-20\%, {\em (d)} 0-20\%; {\em (e)} 0-5\%, {\em (f)} 5-10\%, {\em (g)} 10-15\%, and {\em (g)} 15-20\% most central at 150$A$~GeV/$c$.}
\label{fig:DF2_experiments_NA61}
\end{figure}

Motivated by the positive NA49 Si+Si result, an intermittency analysis was performed on the NA61/SHINE ${}^7$Be~+~${}^9$Be~ \cite{PoS2017} and ${}^{40}$Ar~+~${}^{45}$Sc~\cite{Epiphany2019} systems at 150$A$~GeV/$c$. Results for the correlator $\Delta F_2(M)$ are presented in Fig.\ref{fig:DF2_experiments_NA61}. In the case of Be+Be system, Fig.\ref{fig:DF2_experiments_NA61}(a), $\Delta F_2(M)$ values fluctuate around zero, and no intermittency effect is observed. The result is, however, inconclusive as to criticality, due to the low proton multiplicity of the system.

The Ar+Sc analysis presented here supersedes the older preliminary analysis shown in \cite{Epiphany2019}, as it is based on higher statistics. A scan was performed in centrality, as determined by projectile spectator energy; 5\% and 10\% centrality intervals, as well as the full 0-20\% range, were examined. Selected proton purity was 90\% or higher. Results are shown in \mbox{Fig.\ref{fig:DF2_experiments_NA61}(b-h)}. The scaling effect is weaker, but still consistent, with the earlier analysis; we see a weak indication of intermittency for peripheral collisions \mbox{(10-15\%)}, improving in quality for the wider 10-20\% interval. Reliable $\phi_2$ confidence intervals cannot currently be obtained through simple power-law fits due to bin correlations of $M$ values; the solid red lines in Fig.\ref{fig:DF2_experiments_NA61}(b-h) are simply power-law scaling functions to guide the eye.

\begin{figure}[htb]
\centering
\includegraphics[width=0.45\textwidth]{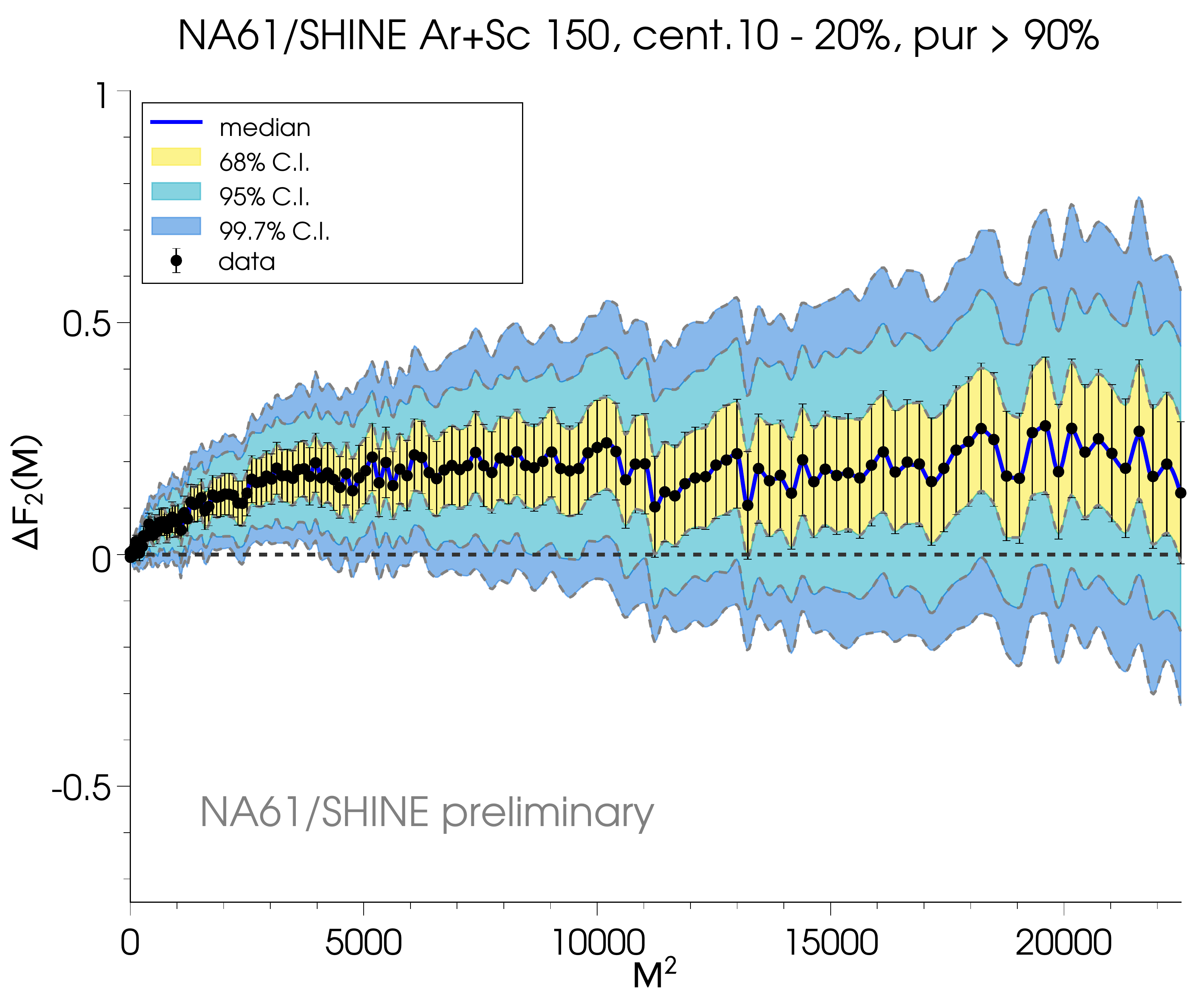}\hspace*{0mm}%
\includegraphics[width=0.45\textwidth]{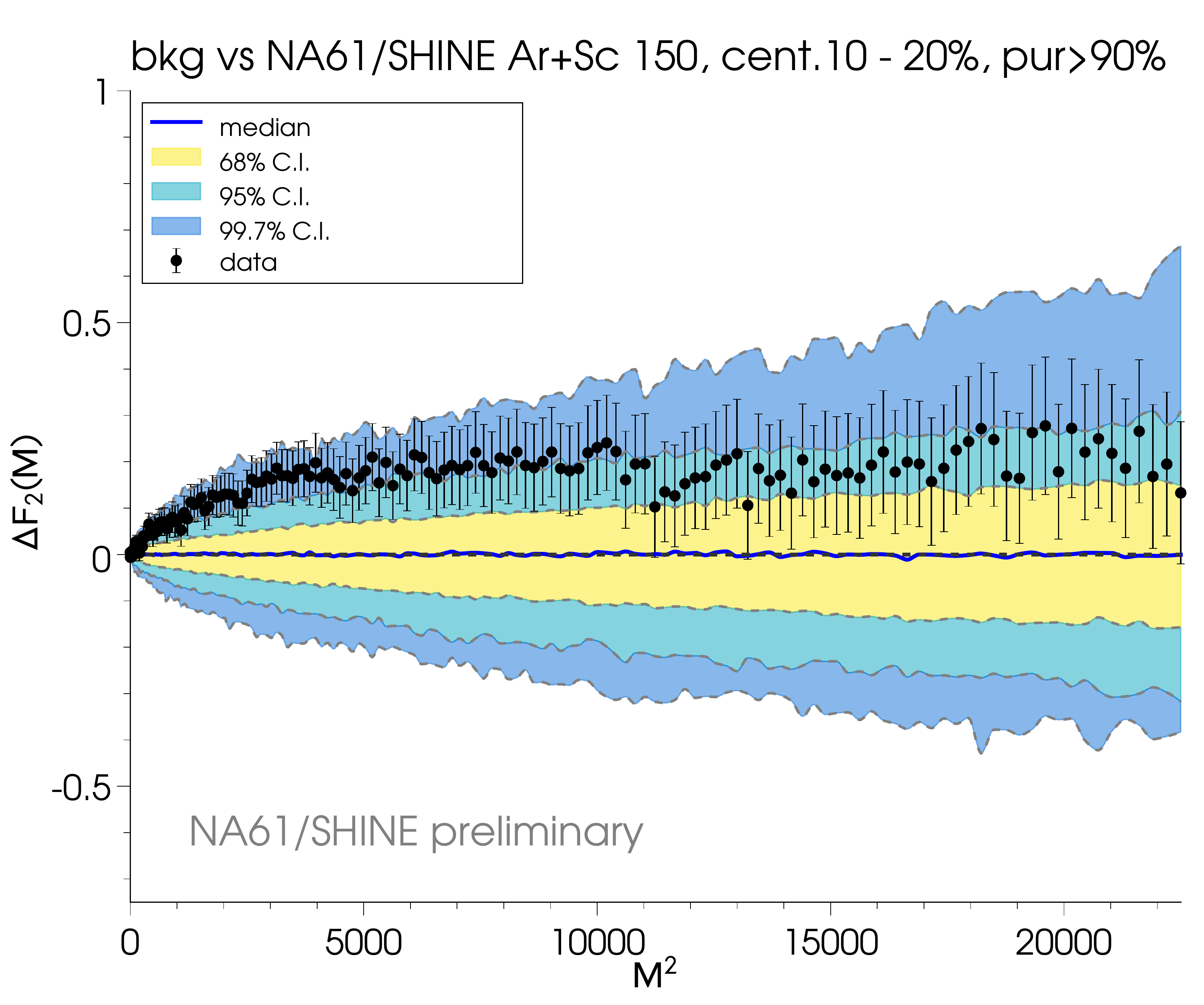}
\caption{{\em Left}: $\Delta F_2(M)$ original sample values for \mbox{10-20\%} central Ar+Sc collisions at 150$A$~GeV/$c$ (black points); error bars correspond to bootstrap standard error; colored bands indicate bootstrap confidence intervals; solid blue line gives the median value of bootstrap samples. {\em Right}: The same experimental $\Delta F_2(M)$ values (black points) compared to the $\Delta F_2(M)$ results for simulated random background protons.}
\label{fig:DF2_significance}
\end{figure}

As mentioned, the uncertainties involved in the $\Delta F_2(M)$ calculation are large; we therefore attempt to quantify the statistical significance of the non-zero effect we see in NA61/SHINE Ar+Sc~@~150$A$~GeV/$c$ data by looking at the bootstrap distributions of $\Delta F_2(M)$ values.

Figure~\ref{fig:DF2_significance}~(\textit{left}) shows the values of $\Delta F_2(M)$ for \mbox{10-20\%} central Ar+Sc collisions at 150$A$~GeV/$c$; original sample data values and their bootstrap standard errors are plotted against confidence intervals (68-95-99.7\%) of the $\Delta F_2(M)$ distributions obtained from 1000 bootstrap samples.\linebreak Figure~\ref{fig:DF2_significance}~(\textit{right}) compares the experimental $\Delta F_2(M)$ values against the $\Delta F_2(M)$ values obtained from an uncorrelated proton background with the same inclusive characteristics as the original Ar+Sc events. Fig.~\ref{fig:DF2_significance} indicates that random background can imitate an effect as large as seen in Ar+Sc in about $\sim5-15\%$ of all cases, and the Ar+Sc effect is above zero in $\sim85-95\%$ of bootstrap samples. Based on these findings, we tentatively assign a 85\% statistical significance to the observed experimental result being not a purely random fluctuation.

\section{Summary and conclusions}

Intermittency analysis of proton density fluctuations in transverse momentum space provides us with a promising set of observables for the detection of the critical point of strongly interacting matter. NA49 Si+Si intermittency analysis at the maximum SPS energy estimates an intermittency index overlapping with the critical QCD prediction, whereas no intermittency is observed in either the smaller C+C or the larger Pb+Pb system at the same collision energy. Preliminary analysis of the NA61/SHINE central Be+Be system at 150$A$~GeV/$c$, consistently, shows no positive result.

We see a first weak indication of a non-trivial intermittency effect in \linebreak NA61/SHINE, in our preliminary analysis of the SSFMs $\Delta F_2(M)$ of Ar+Sc collisions at 150$A$~GeV/$c$. The significance of the effect seems to increase for less central collisions in the case of proton purity thresholds of 90\% and above. However, due to the magnitude of SSFMs uncertainties, and the fact that $F_2(M)$ values for distinct $M$ are correlated, the quality of $\Delta F_2(M)$ power-law scaling remains still to be established, and an estimation of $\phi_2$ confidence intervals is still pending. 

{\small \textbf{Acknowledgments:}  This work was supported by the National Science Centre, Poland (grant no. 2014/14/E/ST2/00018).}


\begin{thebibliography}{99}
%

\bibitem{SHINE} N.~Abgrall \textit{et al.} (NA61/SHINE Collaboration), J.~Inst.~9 (2014) P06005.

\bibitem{Gazdzicki} M.~Gazdzicki, P.~Seyboth, Acta~Phys.~Pol.~B~\textbf{47}, 1201 (2016) and references therein.

\bibitem{Antoniou2005-6} N.~G.~Antoniou \textit{et al.}, Nucl.~Phys.~A~\textbf{761}, 149 (2005); N.~G.~Antoniou \textit{et al.}, Phys.~Rev.~Lett.~\textbf{97}, 032002 (2006). 

\bibitem{Hatta2003} Y.~Hatta and M.~A.~Stephanov, Phys.~Rev.~Lett.~\textbf{91}, 102003 (2003).

\bibitem{Vicsek} T.~Vicsek, \textit{Fractal Growth Phenomena} (World Scientific, Singapore, 1989). ISBN 9971-50-830-3.

\bibitem{Wosiek-Bialas}  J.~Wosiek, Acta~Phys.~Polon.~B~\textbf{19}, 863 (1988); A.~Bia\l{}as and R.~C.~Hwa, Phys.~Lett.~B~\textbf{253}, 436 (1991). 

\bibitem{intermittency} T.~Anticic \textit{et al}, Eur.~Phys.~J.~C~\textbf{75}: 587 (2015).

\bibitem{Metzger} W.~J.~Metzger, {\it ``Estimating the Uncertainties of Factorial Moments"}, HEN-455 (2004) (unpublished).

\bibitem{Efron-Hesterberg} B.~Efron, Ann.~Stat.~\textbf{7}, 1 (1979); T.~Hesterberg \textit{et al.}, \textit{Bootstrap Method and Permutation Tests} (W. H. Freeman \& Co., USA, 2003), ISBN-10:0716757265.


\bibitem{PoS2017} N.~Davis \textit{et al.}, for the NA61/SHINE Collaboration, PoS (CPOD2017) \textbf{054} (2018).

\bibitem{Epiphany2019} N.~Davis \textit{et al.}, for the NA61/SHINE Collaboration, Acta~ Phys.~Polon.~B~\textbf{50}, 1029 (2019).



\end{thebibliography}
\end{document}